%% file: arxiv.tex
\documentclass{article}

\PassOptionsToPackage{numbers}{natbib}

\usepackage[dblblindworkshop, final]{neurips_2026}
\renewcommand{\acksection}{%
  \section*{Acknowledgments and disclosure of funding}%
}

\workshoptitle{TS-LIMITS}

\usepackage[utf8]{inputenc} 
\usepackage[T1]{fontenc}    
\usepackage{hyperref}       
\usepackage{url}            
\usepackage{booktabs}       
\usepackage{amsfonts}       
\usepackage{nicefrac}       
\usepackage{microtype}      
\usepackage{xcolor}         
\usepackage{enumitem}

\usepackage[hang, flushmargin]{footmisc} 
\usepackage{makecell}
\usepackage{multirow}
\usepackage[dvipsnames]{xcolor}
\usepackage{hyperref}
\hypersetup{
    colorlinks=true,
    linkcolor=blue,
    filecolor=magenta,      
    urlcolor=cyan,
    citecolor=blue
}
\usepackage{wrapfig}

\usepackage{comment}
\usepackage{amsmath}
\usepackage[capitalize]{cleveref}
\makeatletter
\AddToHook{cmd/appendix/before}{\def\cref@section@alias{appendix}\def\cref@subsection@alias{appendix}}
\makeatother

\usepackage{mathtools}
\usepackage{algorithm,algorithmicx}
\usepackage[noend]{algpseudocode}
\usepackage[table]{xcolor}

\newcommand{\N}{\mathbb{N}}

\newcommand{\R}{\mathbb{R}}

\newcommand{\tpose}{\text{T}}

\newcommand{\cD}{\mathcal{D}}
\newcommand{\cF}{\mathcal{F}}

\newcommand{\cN}{\mathcal{N}}

\newcommand{\cS}{\mathcal{S}}
\newcommand{\cT}{\mathcal{T}}

\newcommand{\eps}{\varepsilon}

\newcommand{\dequal}{\,\overset{\rm d}{=}\,}

\newcommand{\E}{\mathbb{E}}
\renewcommand{\P}{\text{P}}

\newcommand{\Cov}{\text{Cov}}
\newcommand{\cov}{\text{Cov}}

\newcommand{\simiid}{\overset{\text{\tiny iid}}{\sim}}

\newcommand{\TL}{\cT_{\rm L}}
\newcommand{\TU}{\cT_{\rm U}}
\newcommand{\Trec}{\cT_{\rm rec}}
\newcommand{\Tfit}{\cT_{\rm fit}}
\newcommand{\sizeTL}{T_{\rm L}}
\newcommand{\sizeTU}{T_{\rm U}}
\newcommand{\sizeTrec}{T_{\rm rec}}
\newcommand{\sizeTfit}{T_{\rm fit}}
\usepackage{xspace}
\newcommand{\method}{\text{PPITSAS}\xspace}

\title{Prediction-powered inference for \\ time series across space}

\author{
Shahzar Rizvi$^{1}$\hskip6mm
David Burt$^{2}$\hskip6mm
Vishwak Srinivasan$^{1}$\\
{\bf Renato Berlinghieri}$^{3}$\hskip6mm
{\bf Stefano Del Col}$^{4}$\hskip6mm
{\bf Tamara Broderick}$^{1}$\\[1mm]
$^{1}$MIT LIDS \qquad
$^{2}$UMass Amherst \qquad
$^{3}$Stanford University \qquad
$^{4}$Assicurazioni Generali\\[-0.5mm]
\texttt{shahzar@mit.edu}, \texttt{dburt@umass.edu}, \texttt{vishwaks@mit.edu},\\[-0.5mm]
\texttt{renb@stanford.edu}, \texttt{stefano.delcol@generali.com}, \texttt{tbroderick@mit.edu}
}

\begin{document}
\maketitle
\begin{abstract}
The following motif is common in spatiotemporal settings: we have a sequence of covariate and label pairs observed for a relatively short, recent time period. We have access to unlabeled covariates over a longer time period. Data is observed over many spatial locations. For instance, crop yield might be observed over a large geographical area for recent years, but weather data (which is informative about crop yield) is available for a much longer period. The goal is to estimate, at each spatial location, the expected label (e.g., crop yield) in the future and provide a valid confidence interval for this value. The observed time period alone is too short for reliable estimates. Imputing missing labels with machine learning can cause substantial bias. Prediction-powered inference (PPI) can correct for this bias, but it relies on an i.i.d.\ assumption that breaks under our expected temporal dependencies. Heteroskedasticity and autocorrelation consistent (HAC) procedures account for temporal correlation, but have not been adapted to cases where some labels are imputed. We provide reliable point estimates and confidence intervals given: short labeled time series (across spatial locations), a longer unlabeled time series, and an imperfect predictor of labels given covariates. We show our method outperforms natural alternatives.
\end{abstract}

\section{Introduction}
\label{sec:introduction}

We start with an illustrative motivating example and then describe the general case in more detail. We have been collaborating with Generali Group, one of the world’s largest insurance and asset management groups. Generali faces the following practical risk-modeling setting related to insuring assets. In a range of locations across a wide geographic span, Generali has a historical record of insurance claims that are largely caused by adverse weather. Generali also has access to a longer record of daily weather data, but without corresponding claims information. Generali would like to use this additional information to support risk estimation in locations exposed to weather-related hazards. Moreover, confidence intervals with reliable coverage for these risk estimates are important to support risk assessment and decision-making.

Generali’s problem is an example of a more general motif. In the motif, we have access to data at a set of locations spanning a large geographic area. At each location, we have data across time steps in the recent past consisting of both responses and a collection of covariates that we expect to be informative about those responses. Prior to the labeled data, we have access to a much longer historical sequence of just the covariates (i.e., no responses). We would like to use all the available data to estimate the expected response value in future time periods, and we would like to provide reliable confidence intervals for this estimate.

We expect this motif to arise commonly in practice. For instance, \citet{lupone2024hindcasting} have access to limited recent data labeled with mammal population sizes and longer-term historical data on environmental variables; they build predictors from the shorter-term labeled data and ``hindcast'' mammal populations across the past, unlabeled data. 
Similarly, practitioners are interested in (1) estimating future crop yield using meteorological covariates \citep{fan2022}, and researchers might have started collecting crop yield more recently than the extent of the meteorological record; and (2) estimating future PM\textsubscript{2.5} levels using satellite, meteorological, and land-use covariates \citep{just2015, xiao2022}.

This motif presents challenges to standard inferential frameworks due to the limited labeled data and known temporal dependencies. For instance, a classical and widely used approach to similar problems (without an unlabeled dataset) is to estimate the future response with the empirical average of the observed responses and report a standard confidence interval.
We could just ignore the unlabeled data and apply this approach, but we expect poor point estimations due to the limited availability of labeled data. 
For instance, consider our running insurance example, and suppose a particular location actually experiences a storm every 2 years in the long run. Suppose we have 5 years of labeled data; that, by chance, we only observe one storm in those 5 years; and that an insurance company pays out 10\% of insured asset value (payout = 0.1) when a storm hits. The true mean annual payout then would be $0.1 \times 1/2 = 0.05$ at this location, but the classical estimate would yield $0.1 \times 1/5 = 0.02$.

A natural and widespread correction \citep[e.g.][]{gamazon2015gene,gusev2019transcriptome,vinas2023hypergraph,hou2023surrogate}, which we refer to as ``inference with imputation'' (IWI),\footnote{\citet{wang2020} refer to this method as ``post-prediction inference,’’ but we prefer a name with an acronym distinct from that of prediction-powered inference (PPI).} is to fit a machine learning predictor using some of the available labeled data. IWI uses this predictor to predict the responses from the observed covariates, so all unlabeled points are now assigned (predicted) labels. IWI treats the predicted labels as though they were true labels and applies the classical approach.  In the example above, we might hope that incorporating the historical record of weather data would ensure we have a better estimate of storm frequency relative to the classical approach. However, IWI is known to introduce its own bias \citep{motwani2023revisiting}. For instance, we might expect bias in the machine learning predictions due to unobserved covariates relevant to the problem.

This bias is essentially the problem that prediction-powered inference (PPI) \citep{angelopoulos2023ppi,angelopoulos2024ppiep} 
is designed to solve. PPI uses some of the labeled data to estimate the bias between the predicted and true responses and then uses this estimate to debias the final estimated mean response. While PPI should substantially improve point estimation, we expect its confidence interval to be miscalibrated in our setting since it relies on an assumption that data is i.i.d. In particular, a straightforward way to apply PPI would be to treat the data across time, observed at any single spatial point, as i.i.d. But we know that meaningful dependencies exist across time in our setting.

Another way to frame our problem is the following: PPI relies on an i.i.d.\ assumption that is meaningfully wrong in the spatiotemporal setting of interest to us. In the present work, we propose to work under a more spatiotemporally realistic (and strictly weaker) assumption than i.i.d.\ (namely, time stationarity), and to develop an analogous framework for valid confidence intervals for the long-run mean response of interest.
Concretely, our contributions form two main directions.
\emph{First}, we decide how to efficiently use limited labeled data in the face of temporal correlation. E.g., while the order of data does not matter for standard PPI, we decide between interleaving or chunking data reserved for prediction vs.\ for debiasing.
\emph{Second}, we  adapt heteroskedasticity and autocorrelation consistent (HAC) procedures \citep{whitedomowitz1984dependent, newey1987hac, denhaan1997practitioner, lazarus2018har} within a PPI framework.
PPI confidence intervals rely on estimated uncertainties. HAC corrects uncertainties using estimates of the covariance between response data on nearby time steps. By combining the bias correction of PPI with the uncertainty adaptation of HAC, we aim to construct inferentially valid confidence intervals that leverage machine learning predictions while accounting for the temporal dependence of the observed data. Along the way, we sketch out an asymptotic justification of our approach, relying on our choice of the Newey--West \citep{newey1987hac, newey1994bandwidth} asymptotic regime as opposed to that of \ Kiefer--Vogelsang \citep{kiefervogelsanghac}. Though we have not filled in a full proof, we describe what makes the problem non-trivial and hope to complete the theory in a future version of the manuscript.

In what follows, we start by establishing notation for our problem setup in \cref{sec:setup}. We briefly review past methods in \cref{sec:ppi} along the way to establishing our method, which we call \emph{PPI for time series across space} (\method, pronounced ``pizzas''), in \cref{sec:our_method}. We describe our experimental setup in \cref{sec:experiment_setup}, where we address subtle challenges that arise in validating method performance.  Finally, in \cref{sec:results}, we empirically demonstrate that our proposed method, \method, outperforms the evaluated alternative approaches in our setting.

\subsection{Related work}\label{sec:related-work}
\textbf{Relation to missingness variations.} To adapt PPI to handle temporal dependencies, we choose in the present work to build from the PPI and HAC literatures. 
Alternatively, we might have built from a literature aiming to extend the missingness conditions under which PPI (and similar adjustments) work.
One can conceptualize standard PPI as working under an assumption that responses are ``missing completely at random'' (MCAR) within the full data set. Recent work has focused on extending to ``missing at random'' (MAR), where the missingness indicator can be predicted from the covariates, but not the missing responses themselves. If we include observation times among the covariates, then our setting can be considered a special case of MAR. See \Citet[Section 2]{song2026demystifying} for a discussion of connections between PPI and missingness.

For any method to work, we need a regularity assumption that relates the unlabeled responses to the labeled responses. In MCAR, this is the i.i.d.\ assumption.
\citet{kluger2025} consider two distinct ways to address this need for regularity in MAR.
(1) They make a standard assumption (familiar from the distribution shift literature) that covariates in the missing and not-missing data overlap substantially. If we consider time as a covariate (to fit their framework), our unlabeled data are, by design, strictly prior in time to our labeled data; thus, overlap fails. (2) They remark that more generally, one might extend their approach to cases where  asymptotic normality and bootstrap consistency results hold. Even if a formal result to this effect were available, the work still needs to be done to choose an appropriate regularity assumption (analogous to the i.i.d.\ assumption, distribution shift with overlap, or time stationarity) and work out the asymptotic normality result. We see this work as equivalent in scope to what we accomplish in this paper.

\citet{salerno2026} likewise consider a case with observations missing at random and filled in via prediction, though they more explicitly assume spatial dependence. The distinction between time and space need not be meaningful; we could treat our time coordinate as a spatial coordinate in their framework. However, they again rely on an overlap assumption (to fit propensity models), which again fails when we treat time as a covariate (just as \citet{salerno2026} explicitly treat their spatial coordinates as covariates).

\textbf{Mean estimation vs.\ anytime valid testing.} \citet{csillag2025ppie} demonstrate that any e-value procedure has a PPI counterpart, and give an application in change-point detection for time-series data. \citet{csillag2025ppicp} extend PPI and address a risk monitoring problem with temporal dependence. But both papers address anytime valid sequential testing rather than mean estimation.

\section{Problem setup}
\label{sec:setup}
We next describe the available data and our analysis goals in more detail.

\textbf{Data.} We observe data at a fixed set of locations \(\cS \coloneqq \{s_1,\dots, s_{N_{\cS}}\}\). 
Across all locations, we observe data across the same set of times. For convenience, we assume all times are in the positive integers ($\N =\{1,2,\ldots\}$). E.g., in our running example, each integer corresponds to a day, and we have access to daily weather at many sites across a country for multiple decades. We expect it would be conceptually straightforward, but notationally cumbersome, to extend our development to non-integer times.
We assume the times at which we observe unlabeled data, \(\TU = \{1, \ldots, \sizeTU\}\), occur directly before those with labeled data, \(\TL = \{\sizeTU + 1, \ldots, \sizeTU + \sizeTL\}\).
In particular, we have access to (1) covariates \(X_{t}(s) \in \R^D\) across \(s \in \cS\) and all times \( t \in \TU \cup \TL\), and (2) responses \(Y_{t}(s) \in \R\) across \(s \in \cS\) and \( t \in \TL\).

\textbf{Estimand.} Roughly, our goal is to estimate the long-run mean of the responses at each spatial location. To make this estimand precise, we need some assumptions; the assumptions leading to our estimand follow Section 7.2 of \citep{hamilton1994timeseries}. We start by assuming that, for each spatial location $s$, there exist two stochastic processes indexed by $t \in \N$: the covariate process \(\{X_{t}(s)\}_{t}\) and the response process \(\{Y_{t}(s)\}_{t}\). We assume that, for all $t \in \N$, $\E[Y_t(s)]$ exists, and for all  $t_1, t_2 \in \N$, $\cov[Y_{t_1}(s), Y_{t_2}(s)]$ exists. Finally, we assume that (1)
\(\{Y_{t}(s)\}_{t}\) is covariance-stationary (i.e., for each location $s \in \cS$, there exist real-valued $\mu(s)$ and $\{\gamma_j(s)\}_{j=0}^{\infty}$ such that, for all times $t,t_1,t_2 \in \N$, the left and center of \cref{eq:assumptions-for-estimand} hold) and (2) that the covariances are absolutely summable (right, \cref{eq:assumptions-for-estimand}). 
\begin{equation}
\label{eq:assumptions-for-estimand}
    \E[Y_t(s)] = \mu(s);
    \quad 
    \cov[Y_{t_1}(s), Y_{t_2}(s)] = \gamma_{|t_1 - t_2|}(s);
    \quad
    \sum_{j=0}^\infty | \gamma_j(s)| < \infty.
\end{equation}

With these assumptions in hand, Proposition 7.5 of \citep{hamilton1994timeseries} ensures that, at each spatial location $s$, the long-run mean response ($\theta(s)$ in \cref{eq:long-run-mean}) exists and is well defined:
\begin{equation}
\label{eq:long-run-mean}
    \theta(s) \coloneqq \lim_{T \to \infty} \frac{1}{T} \sum_{t=1}^T Y_t(s).
\end{equation}
Our goal is the following: for each spatial location $s \in \cS$, estimate the long-run mean response and provide a valid confidence interval for it.

\section{Background: Prediction-powered inference}
\label{sec:ppi}
Since \method{} builds on PPI, we first establish the standard PPI method and notation.

To apply PPI, one first partitions the indices $\TL$ of labeled data into indices $\Tfit$ used for fitting the predictor $\smash{\hat{f}}$ and indices $\Trec$ used for debiasing, or ``rectifying,'' the final estimate. Since PPI assumes data and label-missingness are i.i.d., only the sizes $\sizeTfit \coloneqq |\Tfit|, \sizeTrec \coloneqq |\Trec|$ matter, not the order of the indices. We let $\hat{Y}_t(s) \coloneqq \smash{\hat{f}}(X_t(s))$ be the predicted response at location $s$ and time $t$.

For location $s$, PPI constructs its point estimate $\smash{\hat{\theta}^{\rm PPI}(s)}$ by starting from the empirical mean predicted response on the unlabeled data, $\smash{\hat{\mu}(s)}$. It computes an estimate $\smash{\hat{\Delta}(s)}$ of the expected bias $\smash{\Delta(s)}$ of $\smash{\hat{\mu}(s)}$ and subtracts that from $\smash{\hat{\mu}(s)}$:
\begin{align}\label{}
\hat{\mu}(s) &\coloneqq \frac{1}{\sizeTU} \sum_{t \in \TU} \hat{Y}_t(s), &\qquad \hat{\Delta}(s) &\coloneqq \frac{1}{\sizeTrec} \sum_{t \in \Trec} \Big(\hat{Y}_t(s) - Y_t(s)\Big),\label{eq:mu-delta-hat}\\
\Delta(s) &\coloneqq \frac{1}{\sizeTU} \sum_{t \in \TU} \mathbb{E}\Big[ \Big(\hat{Y}_t(s) - Y_t(s)\Big)\Big], &\qquad \hat{\theta}^{\rm PPI}(s) &\coloneqq \hat{\mu}(s) - \hat{\Delta}(s).\label{eq:delta-ppi}
\end{align}

Let $\hat{\sigma}_{\hat{Y}}^{\rm PPI}(s)^2$ be the unbiased sample variance of the predicted responses and $\hat{\sigma}_{\hat{Y}-Y}^{\rm PPI}(s)^2$ be the unbiased sample variance of the bias terms (see \cref{eq:ppi-var-y-hat} and \cref{eq:ppi-var-bias} in \cref{app:previous-methods}). 
If the data is i.i.d., the point estimate $\hat{\theta}^{\rm PPI}(s)$ is the sum of two independent sample means;
so, as discussed in \citep{angelopoulos2023ppi}, an asymptotically valid $(1 - \alpha)$\% confidence interval is 
\begin{equation}
    \label{eq:ppi-estimator}
    \hat{\theta}^{\rm PPI}(s) \pm z_{1-\alpha/2} \sqrt{\hat{\sigma}^{\rm PPI}_{\hat{Y}}(s)^2\big/\sizeTU +\hat{\sigma}^{\rm PPI}_{\hat{Y}-Y}(s)^2\big/\sizeTrec}.
\end{equation}

\section{Our method: PPI for time series across space (\method)}
\label{sec:our_method}

Here, we discuss our proposed method \method for this problem, summarized in \cref{alg:method} in  \cref{app:method}.
We use this section to elucidate design choices that we make in our algorithm.

\subsection{Splitting the data and training the predictor}\label{sec:split-train}

\textbf{Sharing information across space for a better predictor.} As in standard PPI, we partition $\TL$ into disjoint subsets $\Tfit$ and $\Trec$. One might na\"ively apply PPI directly and separately to each spatial coordinate, including fitting a separate predictor to data from each spatial coordinate. However, given our sparsity of labeled data, we instead use data from $\Tfit$ across all locations $s \in \cS$ to fit a single predictor $\smash{\hat{f}}$. With $N_{\cS} \cdot \sizeTfit$ data points to fit $\smash{\hat{f}}$, rather than the na\"ive $\sizeTfit$, we can choose a much more modest $\sizeTfit$---leaving more data for accurate debiasing.

\textbf{Splitting the labeled data.} Unlike in standard PPI, the precise set of indices assigned to $\Tfit$ and $\Trec$ (rather than just their sizes) matters in our setting due to  temporal dependencies in the data. We consider two options: chunking and interleaving. In a chunked split, we preserve the temporal ordering of the labeled data and divide $\TL$ at a single cutoff, using the earliest $\sizeTrec$ time points to form $\Trec$, and the last $\sizeTfit$ time points to form $\Tfit$. W use the later block of $\TL$ for $\Tfit$ so that the predictor is less correlated with $\TU$. In an interleaved split, we set aside observations at evenly spaced times for fitting.

A chunked split exhibits greater correlation within each chunk, compared to interleaving; since the predictor is trained on more contiguous, and hence more correlated data, we expect the chunked split to exhibit more prediction bias ($\smash{\Delta(s)}$ in \cref{eq:delta-ppi}). However, for the chunked split (vs.\ the interleaved split), there is less correlation between the fitted chunk and the rectifying chunk. So we expect a better estimate ($\hat{\Delta}(s)$ in \cref{eq:mu-delta-hat}) of the prediction bias $\Delta(s)$ than using the interleaved split. Finally, then, we expect  that $\hat{\theta}^{\rm PPI}(s)$ in \cref{eq:delta-ppi} applied with chunking should outperform the same estimator applied with interleaving, because its performance depends on the quality of its bias estimation.

\begin{wraptable}{o}{0.52\textwidth}
\centering
\begin{tabular}{lccc}
\toprule
              & {$\Delta$} & {$\E[\hat{\Delta}]$} & {$\E[\hat{\theta}^{\rm PPI} - \theta]$} \\ 
\midrule
Chunking      & $-0.606$ & $-0.602$ & $0.003$ \\
Interleaving  & $-0.080$ & $-0.001$ & $0.078$ \\ 
\bottomrule
\end{tabular}\\[1ex]
\caption{For each row (using a chunked or interleaved split), we report true prediction bias (left column), estimated prediction bias (middle), and bias of the PPI estimate (right). In each column, we report the expected result by averaging over $10^5$ replicates. In every case, the standard error has size strictly less than $3 \times 10^{-3}$; see \cref{tab:bias-demo-full} in \cref{app:demo} for the full standard errors.}
\label{tab:bias-demo}
\end{wraptable}

To test these intuitions, we set up a simple autoregressive model (with a temporal component but no spatial component) in \cref{app:demo}. We let $\sizeTU=4{,}000$, $\sizeTL=1{,}000$, $\sizeTfit=100$. For the interleaved split, we put data points indexed $\{4001,4011,4021,\ldots,4991\}$ in $\Tfit$ and the remaining points in $\Trec$.
We compute expectations and standard errors via Monte Carlo averages across $10^5$ independent replicates. We see in \cref{tab:bias-demo} that the chunking approach has a higher prediction bias, but also a better estimate of that bias---and thus  results in a final estimator $\smash{\hat{\theta}^{\rm PPI}}$ with less overall bias.

\textbf{Buffering.} Even with the chunked split, there are temporally near data at the boundary between the fitting and rectifying sets. We could choose to buffer this data by throwing out some of the observations at the boundary of the two sets, but buffering loses already-scarce labeled data. Alternative options include: (1) do not buffer, and assume that the correlations between the two datasets are negligible and (2) do not buffer but account for, and adjust for, correlations between the two datasets. Presently we choose option (1) based on rough preliminary experiments that suggest negligible changes from buffering; we reserve accounting for induced correlations for future work. 

\subsection{Applying HAC}\label{sec:applying-hac}

Recall that $\smash{\hat{\theta}^{\rm PPI}(s)}$ is the sum of two sample means. So the confidence interval in \cref{eq:ppi-estimator} is asymptotically valid roughly because, under an i.i.d.\ data assumption, (1) the two sample means are independent and (2) $\hat{\sigma}^{\rm PPI}_{\hat{Y}}(s)^2 / \sizeTU$ and $\hat{\sigma}^{\rm PPI}_{\hat{Y}-Y}(s)^2/ \sizeTrec$ are ``good'' estimates of the variance of the respective sample means, $\smash{\hat{\mu}(s)}$ and $\smash{\hat{\Delta}(s)}$. It remains to make a similar argument, and develop an analogous confidence interval, when we no longer have the i.i.d.\ assumption, but instead make a temporal stationarity assumption. We sketch out such an argument here and provide more details, including a detailed discussion of the challenges of a formal proof, in \cref{app:proof-sketch}.

\textbf{Separate treatment of $\smash{\hat{\mu}(s)}$ and $\smash{\hat{\Delta}(s)}$ under temporal dependence.} Under temporal dependence, it will no longer generally hold that $\smash{\hat{\mu}(s)}$ and $\smash{\hat{\Delta}(s)}$ are independent. Instead, we observe that, roughly, under temporal stationarity and standard regularity assumptions discussed in \cref{app:asymptotics}, we can expect $\smash{\hat{\mu}(s)}$ and $\smash{\hat{\Delta}(s)}$ to be asymptotically uncorrelated. If the two are asymptotically normal, this observation will be enough to treat them separately, as in the standard PPI case.

\textbf{Toward a valid confidence interval in the Newey--West regime.} 
HAC methods compute estimated autocorrelations and use them to adjust variance estimates. They have traditionally been applied to the classical estimator we describe in \cref{sec:introduction}. But in the present work we propose to apply HAC to adjust variance estimates for $\smash{\hat{\mu}(s)}$ and $\smash{\hat{\Delta}(s)}$. A parameter called the bandwidth
describes how far (across time) we assume autocorrelations can be non-zero.
There are two major asymptotic regimes to consider. The first is the Newey--West asymptotic regime \citep[Theorem 2]{newey1987hac}, in which the bandwidth is a vanishing proportion of the sample size. The second is the Kiefer--Vogelsang asymptotic regime \citep{kiefervogelsanghac} (often called ``fixed-\(b\)'' asymptotics), in which the bandwidth approaches a fixed proportion of the sample size. 

We first focus on the Newey--West asymptotic limit, where the bandwidth grows more slowly than the length of the labeled and unlabeled time series. In this setting, we conjecture that under reasonable stationarity, mixing, and moment conditions on the response and covariates, as well as a moment condition on the predictor, the following is an asymptotically valid $(1 - \alpha)$\% confidence interval for $\theta(s)$ in our setting:
\begin{equation}\label{eqn:hac-ppi-ci}
   \hat{\theta}^{\rm PPI}(s) \pm z_{1-\alpha/2} \sqrt{\hat{\sigma}^{\rm HAC}_{\hat{Y}}(s)^2\big/\sizeTU +\hat{\sigma}^{\rm HAC}_{Y-\hat{Y}}(s)^2\big/\sizeTrec}.
\end{equation}
We provide a sketch for why this conjecture should be true in \cref{app:proof-sketch}. 
Very roughly, we can apply the Newey--West procedure to find ``good'' (consistent) variance estimates for the individual terms in the sums in $\hat{\mu}(s)$ and $\hat{\Delta}(s)$; call them $\hat{\sigma}^{\rm HAC}_{\hat{Y}}(s)^2$ and $\hat{\sigma}^{\rm HAC}_{\hat{Y}-Y}(s)^2$, respectively (see \cref{eq:hac-var-y-hat} and \cref{eq:hac-var-bias} in \cref{alg:method}). It would then follow from the asymptotic normality of $\smash{\hat{\mu}(s)}$ and $\smash{\hat{\Delta}(s)}$, and their asymptotic uncorrelatedness, that the confidence interval in \cref{eqn:hac-ppi-ci} has the claimed coverage.

\textbf{Newey--West vs.\ Kiefer--Vogelsang.}
In the Kiefer--Vogelsang asymptotics, the limit of the variance estimate is not a constant but a non-degenerate random variable, and so we need not have normality of $\smash{\hat{\mu}(s)}$ and $\smash{\hat{\Delta}(s)}$. Given the resulting added complications of working with Kiefer--Vogelsang and also the standard use of Newey--West in the literature, we henceforth use Newey--West. Additionally, in our motivating example of weather risk and insurance, we have several years of data and expect that the data decorrelates on a scale of several weeks. In this setting where bandwidth is much shorter than the amount of available data, the Newey--West asymptotics should be reasonable. 

\textbf{Newey--West practical choices.} We use the Newey--West procedure with the Bartlett kernel \citep{newey1987hac} and the data-driven plug-in estimate of the bandwidth \citep{newey1994bandwidth}.

\section{Experiments}
After describing our setup, we show that our method (\method) outperforms alternative methods.

\subsection{Experimental setup}
\label{sec:experiment_setup}

\textbf{Necessity of simulated data.} Unlike other common predictive machine learning settings, standard data sets (including Generali's insurance data) do not have ground truth for our setting. To understand why, consider again the thought experiment from \cref{sec:introduction}, where the true (but unknown) mean annual payout is $0.05$, but the empirical mean payout is $0.020$. If we use the observed data to determine ``ground truth'' (by treating the empirical mean as ``ground truth''), a method that estimates the true mean annual payout perfectly will be said to perform ``worse'' (i.e., be farther from the putative ``ground truth'') than methods that underestimate the payout. Simulated data, by contrast, will allow us to access real ground truth.

\textbf{Simulation details.} We aim to create a setup with many similar aspects to the real insurance problem (and other real data problems). See \cref{app:simulation-setup} for full details of our setup. We take a lattice of $N_\cS = 121$ locations. We imagine the scalar covariate $X_t(s)$ as a daily weather measurement and $Y_t(s)$ as a daily payout. We take $\sizeTL = 4{,}000$ labeled times (roughly, 10 years) and $\sizeTU = 16{,}000$ unlabeled times (roughly, 40 years). We generate $X_t(s)$ as a transformation of a Gaussian process across the spatial coordinate. In real life, we generally expect that the predictor cannot fully capture the response's relationship to the covariates (e.g.\ due to unobserved but important covariates). As a particularly simple and interpretable version of this issue, we take $Y_t(s)$ to have a quadratic dependence on $X_t(s)$ and use a linear predictor. We generate versions with more-extreme weather events (heavy-tailed $X_t(s)$ values, which are more realistic for the insurance application) and less-extreme weather events.

\textbf{Performance metrics.} For each method, we report point estimate quality, confidence interval coverage, and confidence interval width. See \cref{app:reporting} for full details. 
For the point estimate, we compute a scaled root mean square error (RMSE); more specifically, for each $s$ and each draw from our data-generating process, we compute the squared error normalized by the (square of) the ground truth $\theta(s)$. We average over these cases and finally take the square root. The scaling allows greater interpretability of the final number; e.g., an RMSE of 1 would suggest errors are on the order of $\theta(s)$ itself in size. Likewise, for confidence interval width, we report the scaled half-width for interpretability, where we again scale by $\theta(s)$. We imagine a practitioner interested in a one-sided hypothesis test---e.g., to test whether the mean response is above a certain threshold. A scaled half-width of 1 suggests the test cannot distinguish $\theta(s)$ from 0.

\textbf{Competing methods.} We compare to the classical method; inference with imputation (IWI); ``chunkPPI,'' a modification of PPI using our chunking approach from \cref{sec:split-train}; and HAC (applied to the classical method). For all methods using a predictor (IWI, chunkPPI, ours), we train the predictor across spatial coordinates as described in \cref{sec:split-train}.

\textbf{Goalposts.} We first look for methods with better point estimation (i.e., lower RMSE). Second, we look for methods that meet near-nominal coverage. Since our nominal coverage is 95\%, we will consider methods to be near-nominal with 85\% or higher coverage. Finally, only for methods with near-nominal coverage, we look for methods to have a lower half-width. We emphasize that half-width is meaningless without coverage; it is trivial to make a confidence interval with arbitrarily small width if we do not require coverage.

To provide a bound on what performance is possible, we also include an oracle comparison; in particular, since we have access to ground truth, we imagine a world where we have responses at all time points (both $\TL$ and $\TU$) and compute the classical estimate and HAC confidence intervals from this (dramatically extended) data---which we do not actually have access to in the problem.

\subsection{Results}\label{sec:results} 
\begin{table}[t]
\centering
\setlength{\extrarowheight}{2pt}
\begin{tabular}{rcccc@{\hspace{0.78em}}ccc}
\toprule
& \multicolumn{3}{c}{\textbf{Less Extreme Weather}}
&
& \multicolumn{3}{c}{\textbf{More Extreme Weather}} \\
\cmidrule(lr){2-4} \cmidrule(lr){6-8}
\textbf{Method}
& \makecell{\textbf{Scaled} \\ \textbf{RMSE}}
& \textbf{Coverage}
& \makecell{\textbf{Scaled} \\ \textbf{Half-Width}}
& 
& \makecell{\textbf{Scaled} \\ \textbf{RMSE}}
& \textbf{Coverage}
& \makecell{\textbf{Scaled} \\ \textbf{Half-Width}} \\
\midrule
Classical
& $0.0503$ & $30.02\%$ & \cellcolor{Gray!25}$0.0194$
&
& $0.0372$ & $31.56\%$ & \cellcolor{Gray!25}$0.0147$ \\

IWI
& $0.0399$ & $21.19\%$ & \cellcolor{Gray!25}$0.0088$
&
& $0.0302$ & $20.79\%$ & \cellcolor{Gray!25}$0.0064$ \\

chunkPPI
& \cellcolor{ForestGreen!25}$0.0299$ & $30.91\%$ & \cellcolor{Gray!25}$0.0115$
&
& \cellcolor{ForestGreen!25}$0.0218$ & $32.88\%$ & \cellcolor{Gray!25}$0.0087$ \\

HAC
& $0.0503$ & $91.72\%$ & $0.0871$
&
& $0.0372$ & $88.00\%$ & $0.0593$ \\

\bf{\method (ours)}
& \cellcolor{ForestGreen!25}$0.0299$
& \cellcolor{ForestGreen!25}$93.45\%$
& \cellcolor{ForestGreen!25}$0.0527$
&
& \cellcolor{ForestGreen!25}$0.0218$
& \cellcolor{ForestGreen!25}$92.64\%$
& \cellcolor{ForestGreen!25}$0.0362$ \\

\midrule
Oracle
& $0.0221$ & $93.72\%$ & $0.0410$
& 
& $0.0166$ & $92.12\%$ & $0.0294$ \\
\bottomrule
\end{tabular}\\[1ex]
\caption{For each method, we report the performance metrics in the less extreme weather setting (left) and the more extreme weather setting (right). The best performance for each metric, according to the goalposts defined in \cref{sec:experiment_setup}, are shaded in green; in the event of a tie, all winners are shaded. We ignore scaled half-widths for coverages below 85\%, and shade them in gray. 
}
\label{tab:results}
\end{table}

\cref{tab:results} displays the performance metrics in the less extreme weather and more extreme weather settings. In both settings, we observe the same patterns. The classical method performs poorly, with the worst scaled RMSEs of all the methods considered. The IWI method improves on the classical method when it comes to point estimation, but this comes at the cost of even worse coverage. The chunkPPI method further improves on point estimation over IWI, with the lowest scaled RMSEs of all methods considered. Our method (\method) achieves the best point estimation of all methods considered; it ties with chunkPPI since they use the same point estimate. In the less extreme weather setting, the scaled RMSE of chunkPPI and \method is about 35\% higher than the scaled RMSE of the oracle. In the more extreme weather setting, the scaled RMSE of chunkPPI and \method is about 31\% higher than the scaled RMSE of the oracle and at least 27\% lower than the scaled RMSE of any of the other methods.

The classical method, IWI, and chunkPPI all fail to achieve coverage close to nominal. As such, their scaled half-widths are not relevant. While HAC performs the same as the classical method in point estimation (since they use the same point estimate), it is able to substantially improve coverage. \method performs even better than HAC, getting the closest to nominal coverage in both settings and essentially tying with the oracle. Between the two methods attaining near-nominal coverages (HAC and \method), \method has the smaller scaled half-widths. Thus \method is uniformly the best in both settings: it has the lowest scaled RMSEs of all of the methods; it has the best coverages; and, among the methods with near-nominal coverages, it has the smallest scaled half-widths. 

\begin{ack}
The authors thank Generali for supporting this work and for their data and insight. VS was supported in part by a MathWorks Fellowship.
RB was supported in part by a Panasonic AI and Wellness Fellowship. The authors thank Tyler McCormick and Stephen Bates for helpful discussions during the early stages of this work. 
Research described in this article was conducted under contract to the Health Effects Institute (HEI), an organization jointly funded by the United States Environmental Protection Agency (EPA) (Assistance Award No.\ CR-84114801) and certain motor vehicle and engine manufacturers. The contents of this article do not necessarily reflect the views of HEI, or its sponsors, nor do they necessarily reflect the views and policies of the EPA or motor vehicle and engine manufacturers.
\end{ack}

\bibliographystyle{unsrtnat}
\bibliography{ref} 

\newpage

\appendix
\section{\method algorithm}\label{app:method}
For the $\mathsf{HAC}$ procedure, see \cref{app:hac}.
\input{neurips-2026/algo-method}
\newpage

\section{Newey--West procedure}\label{app:hac}
\input{neurips-2026/algo-hac}
\newpage

\section{Data splitting demonstration}\label{app:demo}
Consider the following simple autoregressive model, which, for the sake of demonstration, has a temporal dimension only. For $X_1 \sim \cN(0, 1)$, $\{Z_t\}_{t \in \N} \smash{\simiid} \cN(0, 1)$, and $|\rho| <1$, the process is defined for each $t \in \N$ by 
\begin{align}
    X_t &= \rho X_{t - 1} + \sqrt{1 - \rho^2} Z_t\\
    Y_t &= X_t^2.
\end{align}
This process is covariance-stationary with absolutely summable covariances. We aim to estimate 
\begin{equation} 
    \theta \coloneqq \lim_{T \to \infty} \frac{1}{T} \sum_{t=1}^T Y_t.
\end{equation}
Note that $X_t \sim \cN(0, 1)$ for each $t \in \N$. By stationarity, $\theta = \E[Y_1] = \E[X_1^2] = 1$.

We use this example to demonstrate that while interleaving yields a less biased predictor than chunking, it also yields a poorer estimate of the prediction bias. We take $\rho = 0.95$, $\sizeTL=1000$, $\sizeTU=4000$, $\sizeTfit=100$. We have $\TU = \{1, \ldots, 4000\}$ and $\TL = \{4001, \ldots, 5000\}$. 

For the chunking approach, we take $\Trec^{\rm chunk} = \{4001, \ldots, 4900\}$ and $\Tfit^{\rm chunk}= \{4901, \ldots, 5000\}$. For the interleaving approach, we take $\Tfit^{\rm inter} = \{4001, 4011, 4021,\ldots,4991\}$ and $\Trec^{\rm inter} = \TL\setminus \Tfit$. 

\begin{table}[ht]
\centering
\begin{tabular}{rccc}
\toprule
& ${\rm E}[\Delta]$
& ${\rm E}[\hat{\Delta}]$
& ${\rm E}[\hat{\theta}^{\rm PPI} - \theta]$ \\
\midrule
Chunking
& $-0.6058\;(0.0020)$
& $-0.6024\;(0.0022)$
& $-0.0029\;(0.0011)$ \\
Interleaving
& $-0.0796\;(0.0008)$
& $-0.0012\;(0.0002)$
& $-0.0778\;(0.0007)$ \\
\bottomrule
\end{tabular}\\[1ex]
\caption{For each row (using a chunked or interleaved split of the labeled data), we report the true prediction bias (left column), estimated prediction bias (middle column), and bias of the PPI estimate (right column). In each column, we report the result averaged over $10^5$ replicates. Monte Carlo standard errors are reported in parentheses.}
\label{tab:bias-demo-full}
\end{table}

We compute expectations and standard errors via Monte Carlo averages across $10^5$ independent replicates and report them in \cref{tab:bias-demo-full}. As expected, we see that the chunking approach has a higher prediction bias, but also a better estimate of that bias. The PPI estimator $\hat{\theta}^{\rm PPI}$ is less biased when using the chunking approach rather than the interleaving approach.

\subsection{Code and reproducibility}\label{app:data-splitting:code}
The code to generate the simulated data, to run the chunking and interleaving methods, and to compute the reported metrics is available at \url{https://github.com/shahzarrizvi/ppitsas}. The experiment was run locally on a MacBook Pro with an Apple M3 Max chip, consisting of a 14-core CPU (10 performance and 4 efficiency cores), and 36 GB of unified memory. The experiments were CPU-based and did not use GPU acceleration. Generating the simulated data and running the chunking and interleaving methods took less than 5 minutes of computation.

\section{Previous methods}\label{app:previous-methods}
In what follows below, let $s \in \cS$ be the location at which the mean response is being estimated.

\subsection{Classical} 
\label{app:previous-methods:classical}
The classical approach estimates the mean response simply using the sample mean of the observed responses. 
\begin{equation}
    \hat{\theta}^{\rm CLA}(s) \coloneqq \frac{1}{\sizeTL} \sum_{t \in \TL} Y_t(s).
\end{equation}
The confidence interval is constructed by leveraging asymptotic normality. Estimate the variance as
\begin{equation} 
    \hat{\sigma}^{\rm CLA}(s)^2 \coloneqq  \frac{1}{\sizeTL - 1}\sum_{t \in \TL} \left( Y_t(s) - \hat{\theta}^{\rm CLA}(s)\right)^2.
\end{equation}
If \(\{Y_{t}(s)\}_{t \in \sizeTL}\) are independent and identically distributed, then, by the central limit theorem,
\begin{equation}
    \sqrt{\sizeTL} \cdot \frac{\hat{\theta}^{\rm CLA}(s) - \theta(s)}{\hat{\sigma}^{\rm CLA}(s)} \implies \cN(0, 1) \quad \text{ as } \quad \sizeTL \to \infty.
\end{equation}
Under these assumptions, an asymptotically valid confidence interval is 
\begin{equation} 
    \hat{\theta}^{\rm CLA}(s) \pm z_{1 - \alpha/2}\frac{\hat{\sigma}^{\rm CLA}(s)}{\sqrt{\sizeTL}}.
\end{equation}

\subsection{Inference with imputation (IWI)} 

Recall that \(\hat{Y}_{t}(s)\) is the predicted response at location \(s\) and time \(t\).
The IWI estimate is constructed by averaging over the observed responses which were not used for fitting the predictor along with the long sequence of predicted responses.
\begin{align}
        \hat{\theta}^{\rm IWI}(s) \coloneqq \frac{1}{\sizeTrec + \sizeTU}\left(\sum_{t \in  \Trec} Y_t(s) + \sum_{t \in \TU} \hat{Y}_t(s)\right).
\end{align}
When constructing the confidence interval, the predicted responses are treated as true responses. The variance estimate is 
\begin{equation} 
    \hat{\sigma}^{\rm IWI}(s)^2 \coloneqq \frac{1}{\sizeTrec + \sizeTU - 1} \left(\sum_{t \in \Trec} \left(Y_t(s) - \hat{\theta}^{\rm IWI}(s)\right)^2 + \sum_{t \in \TU} \left( \hat{Y}_t(s) - \hat{\theta}^{\rm IWI}(s) \right)^2\right).
\end{equation}
This approach is not generally inferentially valid, even under i.i.d. sampling \citep{wang2020, motwani2023revisiting}. But ignoring this and applying a normal $(1-\alpha)\%$ confidence interval yields
\begin{equation} 
    \hat{\theta}^{\rm IWI}(s) \pm z_{1 - \alpha/2}\frac{\hat{\sigma}^{\rm IWI}(s)}{\sqrt{\sizeTrec + \sizeTU}}.
\end{equation}

\subsection{Prediction-powered inference (PPI)} 
The PPI estimate is constructed by estimating the mean predicted response on the unlabeled data and the mean prediction bias on the labeled data which was not already used for fitting the predictor. 
\begin{equation}
    \hat{\mu}(s) \coloneqq \frac{1}{\sizeTU} \sum_{t \in \TU} \hat{Y}_t(s) \qquad \qquad \hat{\Delta}(s)  \coloneqq \frac{1}{\sizeTrec} \sum_{t \in \Trec} \Big(\hat{Y}_t(s) - Y_t(s)\Big).
\end{equation}
The point estimate $\hat{\theta}^{\rm PPI}(s)$ debiases the average predicted response $\hat{\mu}(s)$ by the average prediction bias $\hat{\Delta}(s)$.
\begin{equation} 
    \hat{\theta}^{\rm PPI}(s) \coloneqq  \hat{\mu}(s) - \hat{\Delta}(s).
\end{equation}
The variances of the predicted responses and the biases are estimated separately.
\begin{align} 
    \hat{\sigma}_{\hat{Y}}^{\rm PPI}(s)^2 &\coloneqq \frac{1}{\sizeTU - 1} \sum_{t \in \TU} \Big(\hat{Y}_t(s) - \hat{\mu}(s) \Big)^2 \label{eq:ppi-var-y-hat}\\
    \hat{\sigma}_{Y - \hat{Y}}^{\rm PPI}(s)^2 &\coloneqq \frac{1}{\sizeTrec - 1} \sum_{t \in \Trec} \Big(\big(\hat{Y}_t(s) - Y_t(s) \big) - \hat{\Delta}(s)\Big)^2 \label{eq:ppi-var-bias}
\end{align}
If the data is independent and identically distributed, the point estimate $\hat{\theta}^{\rm PPI}(s)$ is the sum of two independent sample means; as discussed in \citep{angelopoulos2023ppi}, an asymptotically valid $(1 - \alpha)$\% confidence interval is 
\begin{equation} 
    \hat{\theta}^{\rm PPI}(s) \pm z_{1-\alpha/2} \sqrt{\frac{\hat{\sigma}^{\rm PPI}_{\hat{Y}}(s)^2}{\sizeTU} +\frac{\hat{\sigma}^{\rm PPI}_{Y-\hat{Y}}(s)^2}{\sizeTrec}}.
\end{equation}

\subsection{Heteroskedasticity and autocorrelation consistent procedures (HAC)}\label{app:previous-methods:hac}
This method modifies the classical approach in \cref{app:previous-methods:classical} with a more suitable variance estimate.
The point estimate is the same as \(\hat{\theta}^{\rm CLA}(s)\):
\begin{equation}
    \hat{\theta}^{\rm HAC}(s) \coloneqq \frac{1}{\sizeTL} \sum_{t \in \TL} Y_t(s) = \hat{\theta}^{\rm CLA}(s).
\end{equation}
In lieu of the sample variance, we instead estimate the variance using the HAC variance estimator---summarized in \cref{alg:hac}---as
\begin{equation} 
    \hat{\sigma}^{\rm HAC}(s)^2 = \mathsf{HAC}\Big(\big\{Y_t(s)\big\}_{t \in \TL}\Big).
\end{equation}

Under certain assumptions on the process \(\{Y_{t}(s)\}_{t \in \mathbb{N}}\) (which we elaborate on in \cref{app:hac-asymptotics}), we obtain asymptotic normality of the form
\begin{equation}
    \sqrt{\sizeTL} \cdot \frac{\hat{\theta}^{\rm HAC}(s) - \theta(s)}{\hat{\sigma}^{\rm HAC}(s)} \Rightarrow \cN(0, 1)~.
\end{equation}
Consequently, an asymptotically valid $(1 - \alpha)$\% confidence interval is 
\begin{equation} 
    \hat{\theta}^{\rm HAC}(s) \pm z_{1 - \alpha/2} \frac{\hat{\sigma}^{\rm HAC}(s)}{\sqrt{\sizeTL}}.
\end{equation}
    
\subsection{Oracle}\label{app:previous-methods:oracle}
The oracle method is named as such since it has access to \(\{Y_{t}(s)\}_{t \in \TU}\).
This additional data (which is otherwise unavailable in realistic scenarios) changes the point estimate and the variance estimates.

We have a point estimate:
\begin{equation} 
    \hat{\theta}^{\rm ORA}(s) \coloneqq \frac{1}{\sizeTL + \sizeTU} \sum_{t \in \TL \cup\TU} Y_t(s).
\end{equation}
The variance estimate $\hat{\sigma}^{\rm ORA}(s)^2$ is obtained via the HAC variance estimator (\cref{alg:hac}) as
\begin{equation} 
    \hat{\sigma}^{\rm ORA}(s)^2 = \mathsf{HAC}\Big(\big\{Y_t(s)\big\}_{t \in \TL \cup \TU}\Big).
\end{equation}

Under the same assumptions on \(\{Y_{t}(s)\}_{t \in \TL}\) as in \cref{app:previous-methods:hac}, we have asymptotic normality of the form
\begin{equation} 
    \sqrt{\sizeTL + \sizeTU} \cdot \frac{\hat{\theta}^{\rm ORA}(s) - \theta(s)}{\hat{\sigma}^{\rm ORA}(s)} \implies \cN(0, 1).
\end{equation}
See \cref{app:hac-asymptotics} for more details. This yields the asymptotically valid $(1 - \alpha)$\% confidence interval
\begin{equation} 
    \hat{\theta}^{\rm ORA}(s) \pm z_{1 - \alpha/2} \frac{\hat{\sigma}^{\rm ORA}(s)}{\sqrt{\sizeTL + \sizeTU}}.
\end{equation}

\section{Asymptotics}\label{app:asymptotics}
We do not currently have any complete theoretical results. We conjecture that the HAC confidence intervals in \cref{app:previous-methods:hac,app:previous-methods:oracle} and the PPITSAS confidence intervals in \cref{app:method} are asymptotically valid. We discuss the assumptions we believe are sufficient for the asymptotic validity of these confidence intervals, and sketch out proofs of our conjectured asymptotic validity.
\subsection{Asymptotic validity of HAC confidence intervals}\label{app:hac-asymptotics}
Let $\{W_t\}_{t \in \N}$ be a stochastic process on a probability space $(\Omega, \cF, \P)$, and suppose that for some $\nu > 1$, 
\begin{enumerate}
    \item $\E[|W_1|^{4\nu}] < \infty$,
    \item $\{W_t\}_{t \in \N}$ is $\alpha$-mixing with 
        \[\sum_{j=1}^\infty j^2 \alpha(j)^{(\nu - 1)/\nu} < \infty,\]
    \item $\{W_t\}_{t \in \N}$ is stationary.
\end{enumerate}
We discuss the limitations of these three assumptions in \cref{app:limitations}. 

Condition (1) is a basic moment assumption. Condition (2) is a mixing assumption; in particular, for $\cF_{t}^{t+j} \coloneqq \sigma(W_t, \ldots, W_{t + j})$, we say that $\{W_t\}_{t \in \N}$ is $\alpha$-mixing if the mixing coefficients 
\begin{equation} 
    \alpha(j) \coloneqq \sup_{t \in \N} \sup_{ \substack{A \in \cF_1^t \\ B \in \cF_{t+j}^\infty}} \Big|\P(A \cap B) - \P(A)\P(B)\Big|
\end{equation}
satisfy $\alpha(j) \to 0$ as $j \to \infty$ \citep{bradley2005mixing}. Condition (3) is a stationarity assumption; a process is stationary if for every $m \ge 1$, every $t_1, \ldots, t_m \in \N$, and every $j \in \N$, 
    \[(W_{t_1}, \ldots, W_{t_m}) \dequal (W_{t_1 + j}, \ldots, W_{t_m + j}).\]
That is, finite-dimensional distributions are invariant under time shifts. 

Note that stationarity with the fourth moment condition implies covariance-stationarity, so we have that there exist real-valued $\mu$ and $\{\gamma_j\}_{j=0}^\infty$ such that for all times $t, t_1, t_2 \in \N$, 
\begin{equation}
    \E[W_t] = \mu \qquad \Cov[W_{t_1}, W_{t_2}] = \gamma_{|t_1 - t_2|}.
\end{equation}
By Lemma 1 of \citep{andrews1991hac}, the assumptions also imply that the covariances are absolutely summable:
\begin{equation} 
    \sum_{j=0}^\infty |\gamma_j| < \infty.
\end{equation}
Hence, by Proposition 7.5 of \citep{hamilton1994timeseries}, we have that 
\begin{equation} 
    \theta \coloneqq \lim_{T \to \infty} \frac{1}{T} \sum_{t=1}^T W_t
\end{equation}
exists and is well-defined.

Let $T \in \N$. We show that for 
\begin{equation} 
    \bar{W} \coloneqq \frac{1}{T} \sum_{t=1}^T W_t \qquad \text{ and } \qquad
    \hat{\sigma}^2 \coloneqq \mathsf{HAC}\Big(\big\{W_t\big\}_{t =1}^T\Big),
\end{equation}
the sample mean is asymptotically normal:
\begin{equation} 
    \sqrt{T}\cdot \frac{\bar{W} - \theta}{\hat{\sigma}} \implies \cN(0, 1).
\end{equation}
The stationarity and the fourth moment condition imply that the process is fourth-order stationary. By Lemma 1 of \citep{andrews1991hac}, our assumptions imply that $\{W_t\}_{t \in\N}$ has absolutely summable fourth-order cumulants. Furthermore, by Davydov's covariance inequality \citep{davydov1968covineq, rio1993covineq} with $p = q = 2\nu$, there exists some constant $C > 0$ such that 
\begin{align} 
    |\gamma_j| &\le C \|W_t - \mu\|_{2\nu}  \|W_{t + j} - \mu\|_{2\nu} \alpha(j)^{(\nu - 1)/\nu} &(\text{Davydov})\\
    &= C\|W_t - \mu\|_{2\nu}^2 \alpha(j)^{(\nu - 1)/\nu}. &(\text{stationarity})
\end{align}
Hence for $C' = C\|W_t - \mu\|_{2\nu}^2 < \infty$,
\begin{align} 
    \sum_{j=1}^\infty j^2 |\gamma_j| &\le  \sum_{j=1}^\infty j^2 C\|W_t - \mu\|_{2\nu}^2 \alpha(j)^{(\nu - 1)/\nu}\\
    &= C' \sum_{j=1}^\infty j^2 \alpha(j)^{(\nu - 1)/\nu}\\
    &< \infty.
\end{align}
So our process has $p$-summable autocovariances for $p = 2$. This satisfies the conditions for the consistency of the Newey--West variance estimate $\hat{\sigma}^2$ \citep{newey1994bandwidth}; if 
\begin{equation} 
    s^{(1)} \coloneqq 2\sum_{j=1}^\infty j\gamma_j
\end{equation}
is nonzero, then by Theorem 2 of \citep{newey1994bandwidth},
\begin{equation} 
    \hat{\sigma}^2 \implies \Omega \coloneqq \gamma_0 + 2 \sum_{j=1}^\infty \gamma_j.
\end{equation}
Finally, by Theorem 1.7 of \cite{ibragimov1962clt}, we have that 
\begin{equation} 
    \sqrt{T} (\bar{W} - \theta) \implies \cN(0, \Omega).
\end{equation}
Hence by Slutsky's lemma, we have that 
\begin{equation} 
    \sqrt{T} \cdot \frac{\bar{W} - \theta}{\hat{\sigma}} \implies \cN(0, 1),
\end{equation}
and hence the confidence interval 
\begin{equation} 
    \bar{W} \pm z_{1 - \alpha/2} \frac{\hat{\sigma}}{\sqrt{T}}
\end{equation}
is an asymptotically valid $(1 - \alpha)\%$ confidence interval.

\subsection{Argument for asymptotic validity of \method confidence intervals}\label{app:proof-sketch}
We conjecture that with the following qualitative conditions, the confidence intervals we propose in \cref{eqn:hac-ppi-ci} are asymptotically valid as $\sizeTU, \sizeTL \to \infty$:
\begin{itemize}
    \item Strict-sense stationarity of the covariate and response processes.
    \item Existence of $2+\delta$ moments for the response and predictor.
    \item Sufficiently strong mixing of the covariate and response process.
\end{itemize}
We sketch an incomplete argument for asymptotic validity. First, for a sufficiently strong mixing condition, we should be able to treat a predictor trained on $\Tfit$ as fixed without affecting the rest of the inference process. It suffices that the data predictor is effectively independent from the portions of the processes used to calculate $\hat{\Delta}(s)$ and $\hat{\mu}(s)$. This effective independence of the predictor should be achievable with a suitably strong form of mixing, together with a demonstration that the proportion of data that is correlated with the training data in the rectifying term is asymptotically negligible, or possibly a buffer between the fitting data and the inference data.

Second, we argue that stationarity and a sufficiently strong mixing condition on the covariate and response process $\{(X_t(s), Y_t(s)\}_t$ imply similar conditions on the residual process $\smash{\{\hat{Y}_t(s) - Y_t(s)\}_t}$ and the prediction process $\{\hat{Y}_t(s)\}_t$. 

Third, the mixing, stationarity, and moment conditions imply a joint central limit theorem for $\hat{\mu}(s)$ and $\smash{\hat{\Delta}(s)}$ by \citep[Theorem 1.7]{ibragimov1962clt}. This joint central limit theorem implies asymptotic normality of the PPI estimator, as a linear combination of jointly normal variables is normal. 

Fourth, the mixing condition should imply that as $\sizeTL, \sizeTU \to \infty$, the sample means $\smash{\hat{\Delta}(s)}$ and $\smash{\hat{\mu}(s)}$ become uncorrelated \citep{davydov1968covineq, rio1993covineq}. Therefore, by asymptotic normality, the two terms are asymptotically independent and hence the variance of their sum is the sum of their variances. 

Fifth, \cite{andrews1991hac} demonstrates that stationarity, moment, and mixing conditions yield the Newey--West assumptions for consistent variance estimates when applying the Newey--West procedure \citep{newey1987hac} to $\smash{\{\hat{Y}_t(s)\}_{t \in \Trec}}$ and $\smash{\{\hat{Y}_t(s) - Y_t(s)\}_{t \in \TU}}$, respectively.

Finally, combining the asymptotically consistent estimates of the variances with the joint central limit theorem and Slutsky's lemma yields the asymptotic validity of the confidence interval.

The most subtle part of this argument is the first step, wherein we need to argue that under a sufficient mixing condition, the predictor trained using the same time series data used for inference can be treated as fixed for the remainder of the analysis. Intuitively, this should be true in the asymptotic setting of very long time series, since data at distant times should be effectively independent; therefore the model training cannot affect the estimate of the prediction bias or the estimate of the mean prediction. However, we leave this as a conjecture, and plan to formalize sufficient conditions for this in future work.

\section{Experiments}\label{app:experiments}
\subsection{Simulation setup}\label{app:simulation-setup}
In both settings, we take $\cS = \{0, 0.1, \ldots, 0.9, 1\}^2$ with $N_\cS = 121$. We take $\sizeTL = 4{,}000$ and $\sizeTU = 16{,}000$, corresponding to roughly 10 and 40 years, respectively. From $\TL$, we take $\sizeTfit = 50$ points for fitting a linear predictor and the remaining $\sizeTrec = 3950$ points for rectifying it.

\paragraph{Less extreme weather setting.} 
We model the weather as a linear function added to the Gaussian process and the payout as a noisy measurement of the square of the weather. 

Let $G_t(s) \sim \mathcal{GP}(0, k_{\rm S} \times k_{\rm T})$ be a zero-mean Gaussian process with a covariance kernel which separates into spatial and temporal kernels. We take $k_{\rm S}$ and $k_{\rm T}$ to be Gaussian kernels with length scales $\ell_{\rm S} = 0.1$ and $\ell_{\rm T} = 10$. The weather and payout are then given by 
\begin{align}
    X_t(s) &= s^\tpose \mathbf{1} + 2 + G_t(s)\\
    Y_t(s) &= {\rm clip}_{[0, 1]}\Big((X_t(s)^2 + \eps) / 500 \Big),
\end{align}
where $\eps \sim \cN(0, \sigma^2 I)$ is independent and identically distributed Gaussian noise. We take $\sigma = 0.5$. 

In this case, the weather process $X_t(s)$ has temporal correlations on the order of 1-2 weeks, but is not heavy-tailed. The long-run mean of the payout process $Y_t(s)$ varies across space, but hovers around $0.01$. 

\paragraph{More extreme weather setting.}
We model the weather as a linear function added to the exponential of Gaussian process and the payout as a noisy measurement of the square of the weather. 

As in the less extreme weather setting, let $G_t(s) \sim \mathcal{GP}(0, k_{\rm S} \times k_{\rm T})$ be a zero-mean Gaussian process with a separable covariance kernel governed by length scales $\ell_{\rm S} = 0.1$ and $\ell_{\rm T} = 10$. The weather and payout are given by 
\begin{align}
    X_t(s) &= s^\tpose \mathbf{1} + 10 + \exp(G_t(s))\\
    Y_t(s) &= {\rm clip}_{[0, 1]}\Big((X_t(s)^2 + \eps) / 1200\Big), 
\end{align}
where $\eps \sim \cN(0, \sigma^2 I)$ is independent and identically distributed Gaussian noise. We take $\sigma = 1$. 

In this case, the weather process $X_t(s)$ has temporal correlations on the order of 1-2 weeks and is heavy-tailed. Just like in the less extreme weather setting, the payout process $Y_t(s)$ has a mean which varies across space, but is generally around $0.01$. 

\subsection{Reporting}\label{app:reporting}
In each of the less extreme and more extreme weather settings, we run all of the alternatives (classic, IWI, chunkPPI, HAC, and oracle) and our method \method at a significance level of $\alpha = 0.05$. 

\paragraph{Performance metrics.} In this section, we discuss how the performance metrics are computed.  Let $\smash{\hat{\theta}(s; \cD)}$ and $\smash{\hat{I}(s; \cD)}$ be the point estimate and confidence interval, respectively, corresponding to one of the methods we consider, which takes as input a location $s \in \cS$ and a dataset $\cD$. 

We sample $100$ datasets $\cD_1, \ldots, \cD_{100}$. For the scaled RMSE, we compute the normalized squared errors at each of the $N_{\cS} = 121$ locations. We report
\begin{equation} 
    \sqrt{\frac{1}{121 \cdot 100} \sum_{s \in \cS} \sum_{i=1}^{100} \left(\frac{\hat{\theta}(s; \cD_i) - \theta(s)}{\theta(s)}\right)^2}.
\end{equation}
We compute the estimated coverage of each method. We estimate the coverage by averaging over the 100 sampled datasets and the 121 locations. We report 
\begin{equation} 
    \frac{1}{121 \cdot 100} \sum_{s \in \cS} \sum_{i=1}^{100} \mathbf{1}\big\{\theta(s) \in \hat{I}(s; \cD_i)\big\}.
\end{equation}
We compute the scaled half-width as well. We report
\begin{equation}
    \frac{1}{121 \cdot 100} \sum_{s \in \cS} \sum_{i=1}^{100} \frac{1}{2}|\hat{I}(s; \cD_i)|.
\end{equation}
The resulting statistics are reported in \cref{tab:results}.

\subsection{Code and reproducibility} \label{app:experiments:code}
The code to generate the simulated data for each of the two weather settings, to run each of the methods on the simulated data, and to compute the performance metrics is available at \url{https://github.com/shahzarrizvi/ppitsas}.

These experiments were run locally on a MacBook Pro with an Apple M3 Max chip, consisting of a 14-core CPU (10 performance and 4 efficiency cores), and 36 GB of unified memory. The experiments were CPU-based and did not use GPU acceleration. Generating the simulated data and running the methods on the simulated data for both of the two weather settings took a total of approximately 1 hour of computation.

\section{Limitations}\label{app:limitations}
As discussed in \cref{sec:setup,app:asymptotics}, our method assumes the relevant processes are stationary in time. This assumption is strictly weaker and more realistic than the i.i.d.\ assumption made by standard PPI. In our framework, the stationarity assumption allows for the estimand of the long-run mean response to exist. Stationarity is also what makes the long history of unlabeled data relevant; if the distributions of the covariate and response processes are shifting dramatically in time, we have less reason to believe that leveraging long histories of data will improve our estimation.

However, the assumption can be unrealistic in applications where the data-generating process is changing over time. In our insurance example, changes in the climate or claims practices can result in a mean response which changes over time. We consider developing simulation-based procedures which maintain validity under weaker forms of temporal stability (like local stationarity or smoothly varying distributions) an important direction for future work. 

We also make other regularity assumptions about the covariate and response processes in \cref{app:asymptotics}. By assuming the processes are strongly mixing and satisfy a moment condition, we are able to get a central limit theorem on the sample mean. Without mixing, it is possible for the temporal dependence to persist over arbitrarily long periods of time, making it difficult to get the independence needed for a central limit theorem. The moment condition ensures that sample means, sample variances, and sample autocovariances are sufficiently well-behaved. While such moment conditions are standard, they exclude heavy-tailed processes. In our insurance example, the response process is bounded, and so the moment condition is not a concern.

Another limitation of our present work is in our experiments. We have conducted two experiments comparing the performance of \method against alternatives. In both experiments, we use a linear model to simulate responses from covariates when in actuality they have a quadratic relationship. We intend to consider more realistic simulators (such as a random forest) and more complicated relationships between the covariates and the responses in future work.

\section{Licenses}\label{app:licenses}
In this appendix, we list the libraries and packages used to run our experiments (including versions), and the corresponding licenses.
\begin{itemize}
    \item \verb|numpy==2.0.2|. BSD 3-Clause license.
    \item \verb|scipy==1.13.1|. BSD 3-Clause license.
    \item \verb|scikit-learn==1.6.1|. BSD 3-Clause license.
\end{itemize}
\end{document}

%% file: neurips-2026/algo-method.tex
\algrenewcommand\algorithmicrequire{\textbf{Input:}}
\algrenewcommand\algorithmicensure{\textbf{Output:}}

\begin{algorithm}[ht]
\caption{$\mathsf{PPITSAS}$. PPI for time-series across space}\label{alg:method}
\begin{algorithmic}[1]

\Require Target location \(s\), significance level \(\alpha\), 
fitting proportion \(p_{\rm fit}\), and fitting algorithm \({\sf Alg}_{\rm fit}\).
\Require Labeled data with size $\sizeTL \coloneqq |\TL|$
\[
    \mathcal{D}_{\rm L}
    = \big\{(X_t(s'),Y_t(s'))\big\}_{t \in \TL,\, s'\in \cS}
\]
and unlabeled data with size $\sizeTU \coloneqq |\TU|$
\[
    \mathcal{D}_{\rm U}
    = \big\{X_t(s')\big\}_{\in \TU,\, s' \in \cS}~.
\]

\State Partition \(\TL\) into disjoint, consecutive rectifying and fitting sets
\(\Trec\) and \(\Tfit\) with sizes $\sizeTrec \coloneqq |\Trec|$ and $\sizeTfit \coloneqq |\Tfit|$ such that  $\sizeTfit= p_{\rm fit} \cdot \sizeTL$.

\State Fit the predictor
\[
    \hat f
    \gets
    {\sf Alg}_{\rm fit}
    \Big(
        \big\{(X_t(s'),Y_t(s')) :
        t \in \Tfit,\, s' \in \cS\big\}
    \Big).
\] 
Denote $\hat{Y}_t(s) \coloneqq \hat{f}(X_t(s))$ for each $t$. 

\State Compute the average prediction and bias at \(s\):
\begin{align}
    \hat{\mu}(s)
    &\gets
    \frac{1}{\sizeTU}
    \sum_{t \in \TU}
    \hat{Y}_t(s), \\
    \hat{\Delta}(s)
    &\gets
    \frac{1}{\sizeTrec}
    \sum_{t \in \Trec}
    \Big(
        \hat{Y}_t(s)- Y_t(s) \Big).
\end{align}

\State Estimate the corresponding long-run variances:
\begin{align}
    \hat{\sigma}_{\hat{Y}}^2(s)
    &\gets
    {\sf HAC}
    \left(
        \big\{\hat{Y}_t(s)\big\}_{t \in \TU}
    \right), \label{eq:hac-var-y-hat}\\
    \hat{\sigma}_{\hat{Y} - Y}^2(s)
    &\gets
    {\sf HAC}
    \left(
        \big\{\hat{Y}_t(s)- Y_t(s)\big\}_{t \in \Trec}
    \right).\label{eq:hac-var-bias}
\end{align}

\State Form the \method point estimate
\begin{equation}
\hat{\theta}^{\rm \method}(s)
    \gets
    \hat{\mu}(s)
    -
    \hat{\Delta}(s).
\end{equation}

\Ensure Point estimate \(\hat{\theta}^{\rm \method}(s)\) and
\((1-\alpha)\%\) confidence interval
\begin{equation}
    \hat{\theta}^{\rm \method}(s)
    \pm
    z_{1-\alpha/2}
    \sqrt{
        \frac{\hat{\sigma}_{\hat{Y}}^2(s)}
             {\sizeTU}
        +
        \frac{\hat{\sigma}_{\hat{Y} - Y}^2(s)}
             {\sizeTrec}
    }.
\end{equation}
\end{algorithmic}
\end{algorithm}

%% file: neurips-2026/algo-hac.tex
\algrenewcommand\algorithmicrequire{\textbf{Input:}}
\algrenewcommand\algorithmicensure{\textbf{Output:}}

\begin{algorithm}[H]
\caption{$\mathsf{HAC}$. Heteroskedasticity and autocorrelation consistent variance estimation \citep{newey1987hac}}\label{alg:hac}
\begin{algorithmic}[1]

\Require Time-series observations $\{W_t\}_{t=1}^T$.
\Ensure Newey--West long-run variance estimate $\hat{\sigma}^2$.

\State Compute the sample mean
\begin{equation}
    \bar{W}
    \gets
    \frac{1}{T}\sum_{t=1}^T W_t.
\end{equation}

\State Center the observations:
\begin{equation}
    \xi_t
    \coloneqq
    W_t - \bar{W},
    \qquad t \in \{1, \ldots, T\}
\end{equation}

\State Compute the sample autocovariances
\begin{equation}
    \hat{\gamma}_j
    \gets
    \frac{1}{T}
    \sum_{t=1}^{T-j}
    \xi_t \xi_{t+j},
    \qquad j \in \{0, \ldots, T-1\}.
\end{equation}

\State Set the pilot bandwidth
\begin{equation}
    m
    \gets
    \left\lfloor
        4\left(\frac{T}{100}\right)^{2/9}
    \right\rfloor.
\end{equation}

\State Compute
\begin{equation}
    \hat{s}^{(0)}
    \gets
    \hat{\gamma}_0
    + 2\sum_{j=1}^{m}\hat{\gamma}_j,
    \qquad
    \hat{s}^{(1)}
    \gets
    2\sum_{j=1}^{m}j\hat{\gamma}_j.
\end{equation}

\State Compute the Newey--West bandwidth
\begin{equation}
    \hat{L}
    \gets
    \left\lfloor
        \left(\frac{3}{2}\right)^{1/3}
        T^{1/3}
        \left|
            \frac{\hat{s}^{(1)}}{\hat{s}^{(0)}}
        \right|^{2/3}
    \right\rfloor.
\end{equation}

\State Compute the HAC long-run variance estimate
\begin{equation}
    \hat{\sigma}^2
    \gets
    \hat{\gamma}_0
    +
    2\sum_{j=1}^{\hat{L}}
    \left(
        1-\frac{h}{\hat{L}+1}
    \right)
    \hat{\gamma}_j.
\end{equation}

\State \Return $\hat{\sigma}^2$.

\end{algorithmic}
\end{algorithm}